\documentclass{aa}  

\usepackage{graphicx}
\usepackage{txfonts}
\usepackage[dvipsnames]{xcolor}
\definecolor{comms}{rgb}{0.8, 0.0, 0.0}

\newcommand{\kms}{$\mbox{km\,s}^{-1}$\,}
\newcommand{\nskms}{$\mbox{km\,s}^{-1}$}

\begin{document} 

\title{Clues of the restarting active galactic nucleus activity of Mrk\,1498 from GTC/MEGARA integral field spectroscopy data}

\author{S.~Cazzoli\inst{1} \and L.~Hern{\'a}ndez-Garc{\'i}a\inst{2,3} \and  I.~M{\'a}rquez\inst{1} \and J.~Masegosa\inst{1}  \and G.~Bruni\inst{4} \and 
F.~Panessa\inst{4} \and L.~Bassani\inst{5}}

\institute{IAA-CSIC --  IAA - Instituto de Astrof{\'i}sica de Andaluc{\'i}a (CSIC), Apdo. 3004, 18008, Granada, Spain
\email{sara@iaa.es}
\and
Millennium Institute of Astrophysics, Nuncio Monse\~{n}or S{\'o}tero Sanz 100, Providencia, Santiago, Chile
\and
Instituto de F{\'i}sica y Astronom{\'i}a,  Universidad de Valpara{\'i}so, Av. Gran Breta\~{n}a 1111, Playa Ancha, Chile
\and
INAF -- Istituto di Astrofisica e Planetologia Spaziali, via del Fosso del Cavaliere 100, I-00133 Roma, Italy
\and
Istituto di Astrofisica Spaziale e Fisica Cosmica di Bologna.
}

\date{Received 20 March 2024; accepted 23 August 2024}

\abstract
{Some giant radio galaxies selected at X-rays with active galactic nuclei (AGN) 
show signs of a restarted nuclear activity (old lobes plus a nuclear young radio source probed by giga-hertz peaked spectra). The study of these sources gives us insights into the AGN activity history. More specifically, 
the kinematics and properties of the outflows can be used as a tool to describe the activity of the source.}
{One object in this peculiar class is  Mrk\,1498, a giant low-frequency double radio source that shows extended emission in [O\,III]. This emission is likely related to the history of the nuclear activity of the galaxy. We investigate whether this bubble-like emission might trace  an outflow from either present or past AGN activity.}
{Using a medium-resolution spectroscopy (R\,$\sim$\,10\,000) available with MEGARA/GTC, we derived kinematics and fluxes of the ionised gas from modelling the [O\,III] and H$\beta$ features.  We identified three kinematic components and mapped their kinematics and flux.}
{All the components show an overall blue to red velocity pattern, with similar peak-to-peak velocities but a different velocity dispersion. At a galactocentric distance of $\sim$\,2.3\,kpc, we found a blob with a velocity up to 100\,\nskms, and a high velocity dispersion ($\sim$\,170\,\nskms) that is spatially coincident with the direction of the radio jet. The observed [O\,III]/H$\beta$ line ratio indicates possible ionisation from AGN or shocks nearly everywhere. The clumpy structure visibile in \textit{HST} images at kiloparsec scales show the lowest values of log[O\,III]/H$\beta$ ($<$\,1), which is likely not related to the photoionisation by the AGN.}
{Taking optical and radio activity into account, we propose a scenario of two different ionised gas features over the radio AGN lifecycle of Mrk\,1498. The radio emission suggests at least two main radio activity episodes: an old episode at megaparsec scales (formed during a time span of $\sim$\,100\,Myr), and a new episode from the core ($>$\,2000\,yr ago). At optical wavelengths, we observe clumps and a blob that are likely associated with  fossil outflow. The latter is likely powered by past episodes of the flickering AGN activity that may have occurred between the two main radio phases.}

\keywords{galaxies: active, galaxies: ISM, galaxies: kinematics and dynamics, techniques: spectroscopic.}
\titlerunning{Clues of the restarting AGN activity of Mrk\,1498 from IFS observations}
\authorrunning{S. Cazzoli et al.}
\maketitle


\section{Introduction}
\label{S_intro}

\noindent  Active galactic nuclei (AGN) with radio jets have been observed in phases of activity and dormancy. After a phase of activity ($\sim$\,10$^{8}$\,yr; \citealt{Schawinski2015}), the fuelling of the radio jets stops and the AGN enters a dormant phase.  In some cases ($\sim$\,15\,$\%$), a new phase of restarted activity can be detected even before the older phase has faded \citep{Kukreti2023}. \\
\noindent  Giant radio galaxies (i.e. radio galaxies with a linear extent above 0.7\,Mpc; \citealt{Garofalo2022}) with old lobes that represent a historical record of past activity are perfect laboratories for studying intermittent activity and galaxy evolution. They hold huge reservoirs of energy that can be observed even when it is not fed by the jets \citep{Dabhade2020}. Hence, the same dataset includes different phases of nuclear activity from the same nucleus. \citet{Bassani2016} performed radio and X-ray studies of the combined  INTErnational Gamma-Ray Astrophysics Laboratory (INTEGRAL) and Swift AGN populations from the 3CR catalogue, which includes 67 sources, the linear sizes of 22\,$\%$ of which exceed 0.7 Mpc. These are thus classified as giant radio galaxies.  \citet{Bruni2019} presented a detailed radio study of 13 of these sources and remarkably, 8 sources (61\,$\%$) showed signs of restarting activity in a gigahertz-peaked spectrum (GPS).\\

\noindent  The nuclear activity of galaxies is also modulated by the onset of large-scale outflows. Hence, a complementary view to radio activity that can be used to date the AGN life cycle is to search for evidence for AGN-driven outflows.  These latter have been detected through spectroscopy (lines with blue/red wings or large widths; e.g. \citealt{Cazzoli2014, Cazzoli2018, Cazzoli2020, Cazzoli2022, Maiolino2017}) and imaging (spatially extended or bubble-like emission; \citealt{Masegosa2011, LHM2022}). The outflow properties (e.g. kinematics, lifetime, and energy) can be used to investigate the history of AGN activity. The outflow properties themselves sensitively depend on the history of the AGN energy injection (e.g. the crossing timescale of the outflow), that is, they are sensitive either to faded or restarted  AGN phases \citep{Zubovas2020}. It is still not well understood whether the effect of feedback evolves with the AGN lifecycle.\\
  
\noindent  Interestingly, the optical morphology of giant radio galaxies reveals some potential candidates for diffuse emission that might arise from AGN-driven outflows and hence might be related to the restarted or fading radio activity. In particular, it was shown that the radio activity can also drive outflows (jet-driven outflows; see e.g. \citealt{Tadhunter2007}, \citealt{Hardcastle2020, Morganti2021, Cazzoli2022, Cresci2023, Speranza2024}).\\
The nearby AGN Mrk\,1498 (also called WN 1626+5153, 1626+518 or Swift J1628.1+514; \textit{z}\,$\sim$\,0.055) is a giant radio galaxy and of the best candidates for studying the relation between faded and restarted AGN activity and diffuse gaseous emission. It is a Fanaroff-Riley\,II radio source showing a classical double-lobed structure (projected separation of 1.1\,Mpc) and an unresolved nucleus found by \citet{Rottgering1996} in the Westerbork radio telescope data, more specifically, in data from the Westerbork Northern Sky Survey (WENSS) at 325 MHz. Deep LOFAR images show no connection between the nucleus and the radio lobes \citep{Bruni2021}. At optical wavelengths,  Mrk\,1498 has a prominent broad-line region, as suggested by the very broad H$\alpha$ component (FWHM\,$\sim$\,6500\,\nskms), which resulted in a Sy1.9 classification \citep{Winter2010}. Moreover, Mrk\,1498 shows remarkable extended features that are especially evident in Hubble Space Telescope (\textit{HST}) [O\,III] images \citep{Keel2015, LHG2019}. These features are relatively [OIII] bright bubble-like structures that extend out to $\sim$\,10$\arcsec$ ($\sim$\,10\,kpc) to the north-east and south-west. They are not aligned with the 1.4 GHz radio jet emission (from the NRAO VLA Sky Survey source catalogue; \citealt{Keel2012}) or with the soft X-ray emission from the narrow-line region (NLR), as suggested for type 1 AGNs \citep{Gomez2017}. This misalignment and the distance of the bubble-like diffuse [O\,III] emission rule out any possible relation with the NLR, making its origin intriguing.\\
Taking into account the multi-wavelength properties of  Mrk\,1498, previous studies suggested two main possible scenarios to explain the origin of the diffuse bubble-like emission: It might be the aftermath of a merging event or an AGN-driven outflow. The latter is more likely due to the morphology of the features.\\
Using optical integral field spectroscopy (IFS) observations with the Gemini Multiple-Object Spectrometer (GMOS), \citet{Keel2017} demonstrated that a circumnuclear [O\,III] ring follows the main rotation of the galaxy.  These authors also suggested that the clouds in the extended emission are most likely ionised by either a faded AGN or the aftermath of a past merger episode (see also \citealt{Sartori2016}).\\

\noindent  In this paper, we use the spectral (R\,$\sim$\,10\,000) capability of  MEGARA/GTC optical IFS observations to study the [O\,III]-bright emission in Mrk\,1498.  Our goal is to obtain clues about the AGN activity in Mrk\,1498 by trying to connect the AGN lifecycle to feedback by searching for outflow signatures. We investigate the multiple distinct kinematic components and explore line flux ratios to constrain the dominant ionisation mechanisms.  Furthermore, we make use of the radio properties of Mrk\,1498 to link the AGN restarting and fading activity to the observed optical properties.\\

\noindent  This paper is organised as follows. In Section\,\ref{S_datared}, the data and observations are presented, as is the data reduction. In Section\,\ref{S_analysis}, we describe the spectroscopic analysis based on line modelling and map generations. Section\,\ref{S_results} highlights the main observational results, which are discussed in Section\,\ref{S_discussion}. The main conclusions are presented in Section\,\ref{S_conclusions}.\\

\noindent All images and spectral maps are oriented following the standard criterion, that is, north is up, and east is to the left. Throughout the whole paper, angular dimensions are converted into physical distances using the scale distance from the Local Group (1188 pc/$\arcsec$; see Table\,\ref{T_properties}).


\begin{table}
\caption[Properties]{General properties of  Mrk\,1498}
\begin{center}
\tiny{\begin{tabular}{l  c  c}
\hline \hline
Properties                & value                                 &  References\\
\hline
R.A. (J\,2000)            & 16$^{\rm h}$28$^{\rm m}$04$^{\rm s}$  & NED \\
Decl. (J\,2000)           & +51$^{\rm d}$46$^{\rm m}$31$^{\rm s}$ & NED \\
\textit{z}                         &  0.055625                             & \citet{Markarian1983}\\
D [Mpc]                   &  241.99\,$\pm$\,16.99                             & NED \\
Scale  [pc/$\arcsec$]     &  1188                                  & NED \\
Morphology                & E                                     & \citet{LHG2019}  \\  
\textit{i}\,[$^{\circ}$]  & 38.4                                  & Hyperleda \\
Nuclear Spectral Class.   & Sy\,1.9                               & \citet{Winter2010} \\	
PA$_{\rm jet}$\,[$^{\circ}$]            & 45                                   & \citet{LHG2019}  \\  
\hline
\end{tabular}
\label{T_properties}}
\end{center} 
\tiny{Notes. ---   \lq Distance\rq \ and \lq Scale\rq \ are  from the Local Group.  \lq Morphology\rq: Hubble classification.  \lq \textit{i} \rq \ is the inclination  angle defined as the angle between  the line of sight and the polar axis of the galaxy. It is determined from the axis ratio of the isophote in the B band using a correction for intrinsic thickness based on the morphological type. \lq PA$_{\rm jet}$\rq \ is measured from radio images at different spatial scales. Specifically, with images generated from  NVSS (arcminute scale) and VLBA (arcsecond scale) data obtained at frequencies of 1.4 and 4.8 GHz, respectively.} 
\end{table}


\section{Observations and data reduction}
\label{S_datared}

\noindent Observations were carried out on 2019 July 27 with MEGARA (Multi-Espectr{\'o}grafo en GTC de Alta Resoluci{\'o}n para Astronom{\'i}a; \citealt{Carrasco2018, GilDePaz2018, GilDePaz2020}) the integral field unit (IFU) at the 10.4 m Gran Telescopio Canarias (GTC). The MEGARA IFU consists of 567  fibres (100\,$\mu$m in core size) arranged on a square microlens array that projects a field of 12$\farcs$5\,$\times$\,11$\farcs$3 on the sky. Each microlens is a hexagon inscribed in a circle  with a diameter of 0$\farcs$62 projected on the sky. A total of 56 ancillary  fibres (organised in bundles of eight fibres), located  at a distance of $\sim$\,2.0\,arcmin from the centre of the IFU field of view (FoV) deliver simultaneous sky observations.\\
 We used the  MR-R volume-phased holographic (VPH; i.e. VPH521-MR) covering the 4963\,--\,5445\,\AA \ spectral range  with a spectral resolution of R\,$\sim$\,12\,000. The linear dispersion was $\sim$\,0.122\,\AA/pixel, hence $\sim$\,7.3\,km\,s$^{-1}$ at the corresponding wavelength of [O\,III]$\lambda$5007.  We obtained six exposures of 1360\,s each to facilitate cosmic-ray removal.  During the observations of Mrk\,1498, the dimm-seeing was 1\arcsec \ and the airmass was 1.15.\\
 
\noindent MEGARA raw  data were reduced with the data reduction package provided by Universidad Complutense de Madrid  (\textsc{megara drp}\footnote{\url{https://github.com/guaix-ucm/megaradrp/}}; version 0.8; \citealt{Pascual2019,Pascual2018})  following the MEGARA cookbook\footnote{\url{https://doi.org/10.5281/zenodo.1974954}}. The pipeline allowed us to perform the following steps: sky and bias subtraction, flat-field correction, spectrum tracing and extraction, correction for fibres and pixel transmission, and wavelength and flux calibration (see e.g. \citealt{Cazzoli2020, Cazzoli2022}). \\ 
We  applied a regularization grid to obtain square spaxels\footnote{Hereafter, we call  \lq spaxels\rq all the spatial elements (pixel) in the cube that were obtained after the application of the regularization grid (not those from the fully reduced RSS cube).} with a size of 0$\farcs$4. The final cube has dimensions of 33\,$\times$\,30\,$\times$\,4300, which is equivalent to a total of 990 spectra in the datacube.\\

\begin{figure}
\centering
\includegraphics[clip=true,width=.49\textwidth]{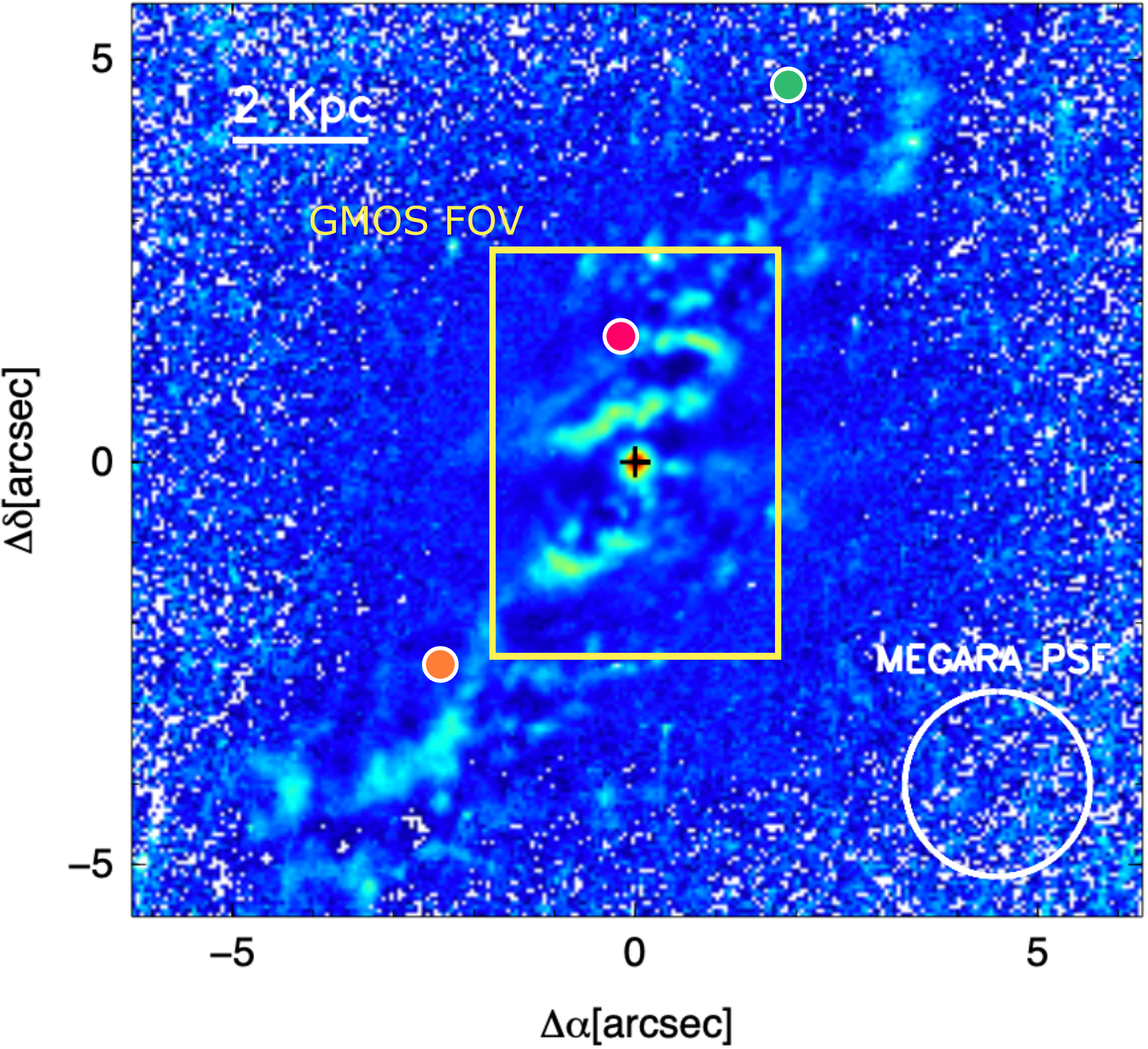} 
\caption{\textit{HST} sharp-divided image obtained from the data of the Advanced Camera for Surveys using the FR551N filter. It was obtained following the receipt in \citet{Marquez2003}, i.e. by dividing the original image, I, by a filtered version of it.  The image is displayed with a zoomed-in view that matches the FoV of our MEGARA observations (Sect.\,\ref{S_datared}). The cross is the photometric centre. The size of the MEGARA PSF is indicated in the bottom right part of the figure.  The white bar in the upper left corner represents 2\,kpc  (1$\farcs$7). The green, pink, and orange circles mark the spatial location at which we extracted the line profiles (and the corresponding modelling) shown in Fig.\,\ref{Fig_ajustes}. The yellow box marks the GEMINI/GMOS field of view from previous IFS observations \citep{Keel2017}.}
\label{Fig_HST}
\vspace{0.5cm}
\end{figure}

\noindent The point spread function (PSF) of the MEGARA datacube can be described as a Moffat function \citep{Moffat1969}. In order to avoid any possible PSF contamination in the kinematic measurements, we conservatively considered as the \lq nuclear region\rq \ a circular area with a radius equal to the width at 10\,$\%$ intensity of the PSF radial profile. This corresponds to 1$\farcs$2 (in radius), as measured from the 2D profile brightness distribution of the standard star. This area is marked (with a circle) in the images and spectral maps from the MEGARA cube and also in the \textit{HST} image in Fig.\,\ref{Fig_HST}  The continuum emission from GTC/MEGARA observations (see Fig.\,\ref{Fig_HST}) covers nearly the entire extension of the diffuse [O\,III]-bright emission. \\
 
\noindent In each spectrum (i.e. on a spaxel-by-spaxel basis), the effect of instrumental dispersion ($\sigma_{\rm\,INS}$, i.e. $\sim$\,0.3\,\AA) was corrected for by subtracting it in quadrature from the observed line dispersion ($\sigma_{\rm\,obs}$): $\sigma_{\rm\,line}$\,=\,$\sqrt{\sigma_{\rm\,obs}^{2}\,-\,\sigma_{\rm\,INS}^{2}}$.\\


\section{Data analysis}
\label{S_analysis}
\noindent The spectral coverage of VPH521-MR is limited to $\sim$\,500\,\AA \ (Sect.\,\ref{S_datared}) and the observed spectra lack the stellar features required for an optimal modelling of the stellar continuum. Hence, we did not apply any procedure to subtract the underlying stellar light. The stellar contribution to the Balmer lines in the nuclear spectrum of Mrk\,1498 is low (see Fig.\,4 in \citealt{LHG2019}).\\

\begin{figure*}
\centering
\includegraphics[width=.95\textwidth]{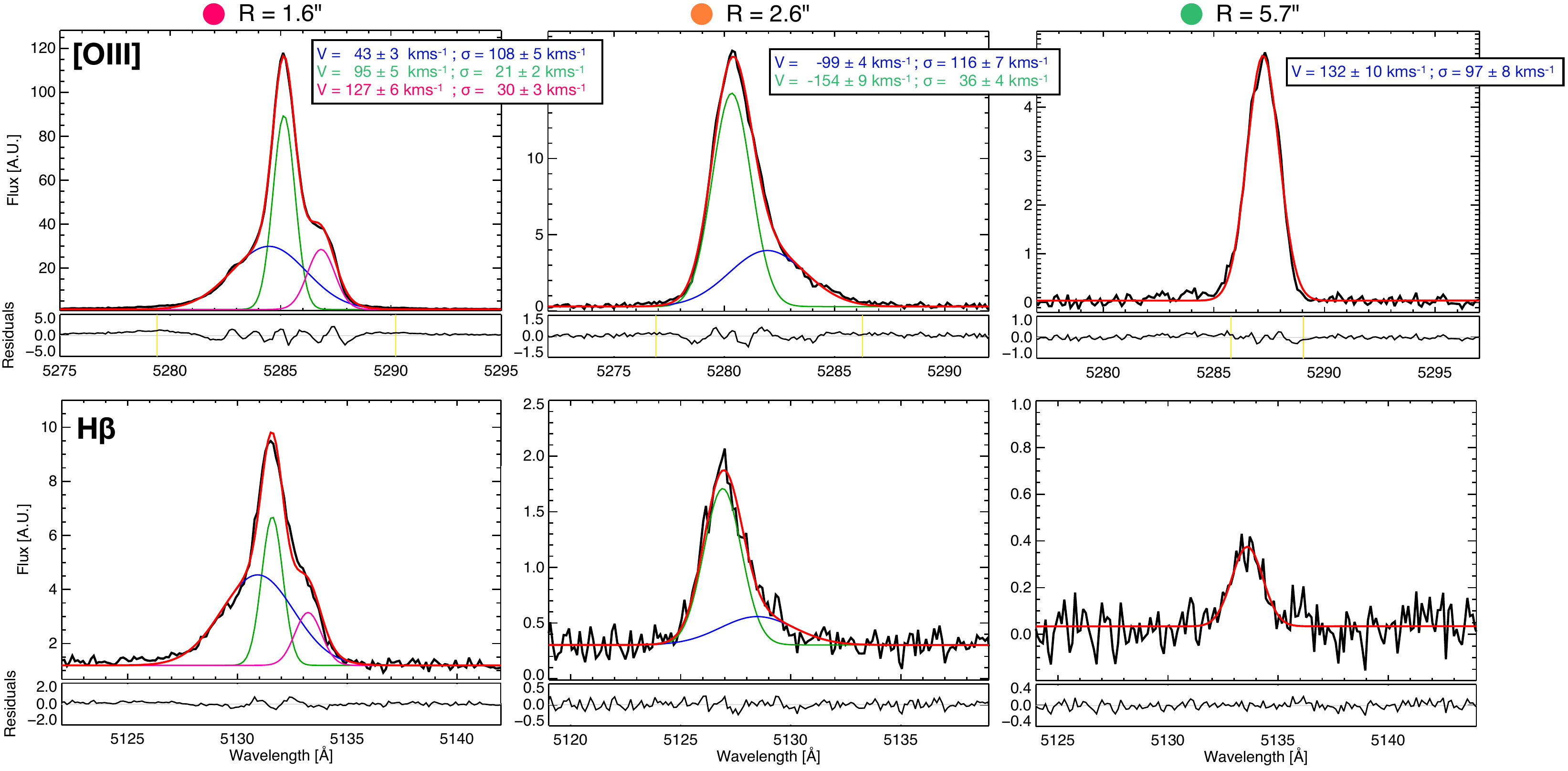}
\caption{Examples of emission line spectra (black) and their modelling at different galactocentric distances (R; see the top labels and symbols in Fig.\,\ref{Fig_HST}). Both [O\,III]$\lambda$5007 (top) and H$\beta$ features are displayed at the corresponding observed wavelengths (i.e. no correction for redshift; see Sect.\,\ref{S_analysis}). For each panel, we show the modelled line profile (red line) and the components (with different colours).  Specifically, Gaussian curves indicate the primary (green), secondary (blue), and tertiary (pink) components we used to model the profiles. Residuals from the fit are shown in the small lower panels, in which vertical yellow lines mark the wavelength range we considered to calculate the $\varepsilon_{\rm line}$ (for the [O\,III]$\lambda$5007 line only; ee Sect.\,\ref{S_analysis} for details). The insets in the upper panels indicate the kinematic values of the components (the emission lines were tied to share the same kinematics).} 
\label{Fig_ajustes}
\end{figure*} 

\begin{figure*}
\centering
\includegraphics[width=1.0\textwidth]{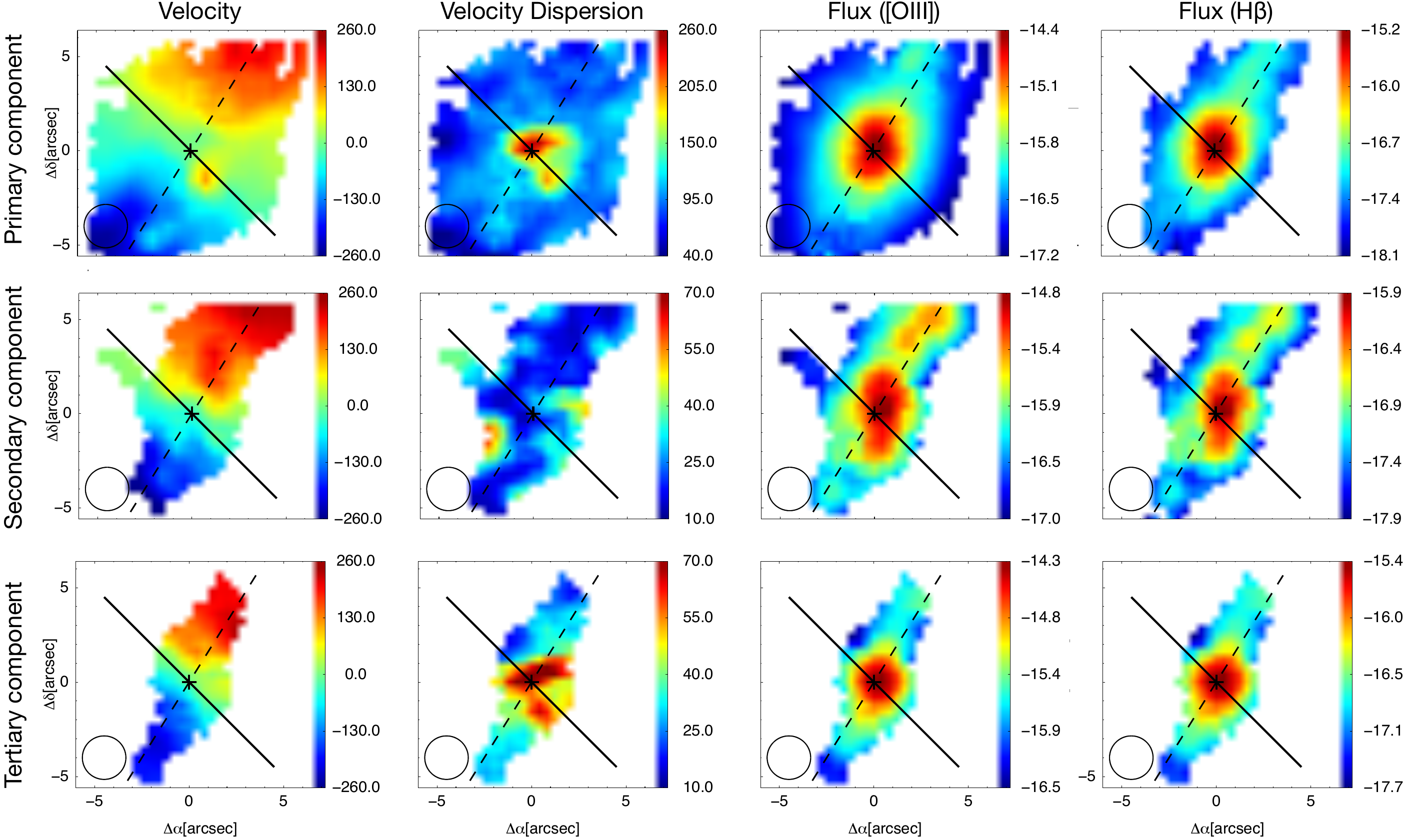}
\caption{[O\,III]$\lambda$5007 velocity field (\nskms) and velocity dispersion (\nskms) maps and flux intensity maps (erg\,s$^{-1}$\,cm$^{-2}$) for the [O\,III]$\lambda$5007 and H$\beta$ emission lines  for the primary component, the broadest of the detected components (Sect.\,\ref{S_analysis}). The cross marks the position of the photometric centre. The solid black line indicates the orientation of the radio jet (PA$_{\rm jet}$\,=\,45$^{\circ}$; Table\,\ref{T_properties}). The dashed lines indicate the PA we used to extract the diagrams shown in Fig.\,\ref{Fig_PV_PS_MajAx}. The maps are spatially smoothed (boxcar average) with a smoothing window of two spaxels (i.e. 0$\farcs$8). The size of the MEGARA PSF is indicated with a circle (as in Fig.\ref{Fig_HST} and in all the maps presented in this paper).} 
\label{Fig_maps}
\vspace{0.5cm}
\end{figure*} 

\begin{figure*}
\centering
\includegraphics[clip=true,width=1.0\textwidth]{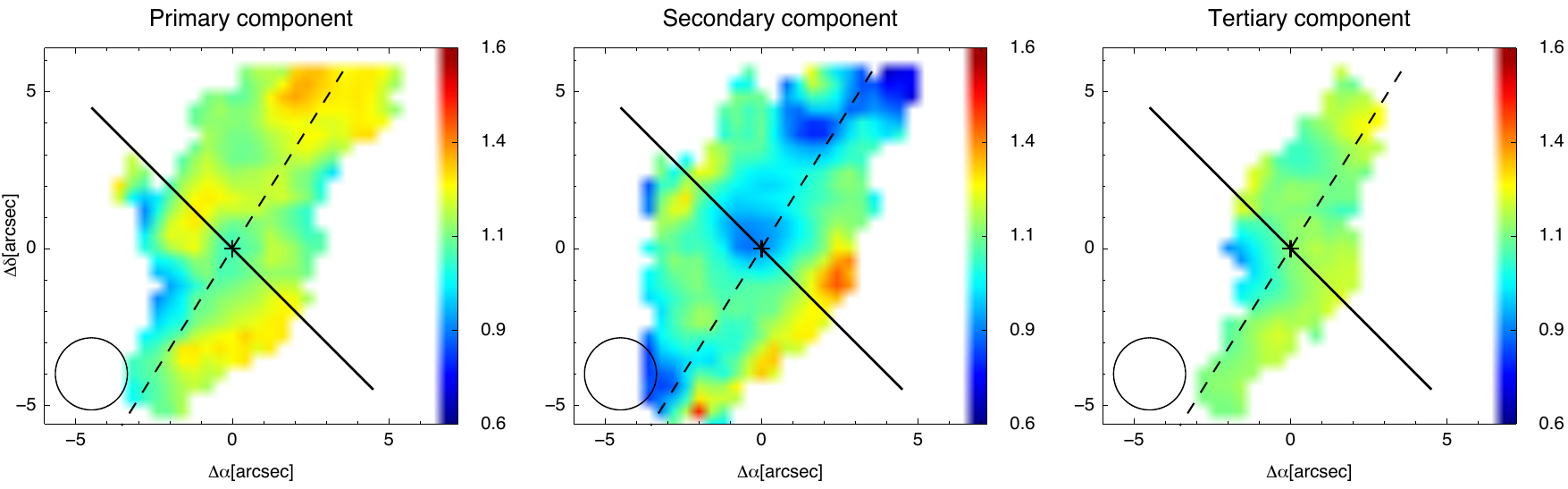} 
\caption{Maps of the standard BPT line ratio log\,[O\,III]/H$\beta$ for the different components. The maps are smoothed as in Fig.\,\ref{Fig_maps}. The symbols are the same as in Fig.\,\ref{Fig_maps}.} 
\label{Fig_map_bpt}
\vspace{0.5cm}
\end{figure*} 


\noindent The spectra in the data cube were visually inspected in order to search for line asymmetries, such as broad wings or double peaks. The latter are present at nearly any distance from the nucleus\footnote{Double-peak emission line signatures detected outside the centre of galaxies indicate preferentially peculiar kinematics (i.e. multiple line-of-sight velocities) and not dual AGNs.} in the H$\beta$ and [O\,III] emission lines. Hence, we modelled the spectral lines with multiple Gaussian functions (up to three per line; see below). Figure\,\ref{Fig_ajustes} shows examples of the line modelling.\\

\noindent All the spectra in the cube were shifted to the rest-frame wavelength using the value of the redshift of Mrk\,1498  (\textit{z}\,=\,0.055625; Table\,\ref{T_properties}). For the line fitting, we then applied a Levenberg-Marquardt least-squares fitting routine under the Interactive Data Analysis (IDL), using \textsc{mpfitexpr} by \citet{Markwardt2009}. Following \citet{Osterbrock2006}, we imposed the intensity ratios of the [O\,III]$\lambda$4959,5007 lines to be 2.99. Moreover, H$\beta$ and [O\,III] were constrained to share the same kinematics (line width and shifts).\\

\noindent  We excluded from the fitting procedure spectra with a signal-to-noise ratio (S/N) per pixel in [O\,III]$\lambda$5007 lower than 3. This excluded 31\,$\%$ of the total number of spaxels.\\
The [O\,III]$\lambda$5007 lines are more intense than H$\beta$ (see e.g. Fig.\,\ref{Fig_ajustes}). In 27\,$\%$ of the total number of spaxels, H$\beta$ was not detected, and we only modelled the [O\,III] lines. These  spaxels are generally located at relatively large distances from the nucleus (i.e. $>$\,4$\farcs$0).\\

\noindent We identified up to three kinematic components to account for double peaks and/or broad wings in the line profiles. To perform the fitting and in order to prevent overfitting, we followed the approach proposed by \citet{Cazzoli2018}, which was successfully applied to optical spectra of active galaxies both from long-slit \citep{LHG2019, LHM2020} and IFS \citep{Cazzoli2020, Cazzoli2022}. Briefly, after fitting all emission lines with one Gaussian component, we evaluated the amplitude of the residuals below the [O\,III]$\lambda$5007 emission line using the $\varepsilon_{\rm line}$ parameter. This parameter is defined as the ratio of the standard deviation of the residuals below the emission lines and that of the line-free continuum. If $\varepsilon_{\rm line}$ was greater than  2.5, a further component was added.  This evaluation was then repeated after we tested the modelling with two components, and a  tertiary component was added if needed. Three Gaussian per line is the largest meaningful number of components that ensures a trade-off between the recovery of emission line properties, the S/N, and the goodness of the fit. \\
 Finally, all the fitting were inspected by eye to ensure a satisfactory line modelling.  Figure\,\ref{Fig_ajustes} shows the results of the line fitting for three regions for which three, two, and one Gaussian component were required, respectively.

\noindent Our final modelling of H$\beta$ did not require a broad component in the nuclear region to confirm the type 1.9 AGN classification (Table\,\ref{T_properties}) of the active nucleus in Mrk\,1498.

\noindent The three components can clearly be distinguished based on their spatial distribution and widths. Specifically, the primary component is detected in near all the FoV, with line widths that are generally larger than 40\,\kms (with values up to 250\,\nskms). This provides a good fit of the line wings. The other two components are mostly observed  in the north-south direction and have narrower widths, $\sigma$\,$<$\,60\,\kms (secondary) and $\sigma$\,$<$\,80\,\kms (tertiary; see Figure\,\ref{Fig_maps}).\\

\noindent We received the following information from the Gaussian functions we used to model each emission line and component: central wavelength, width, and flux intensity, along with their respective fitting errors. These are the 1-$\sigma$ parameter uncertainties weighted for the reduced $\chi^{2}$ of the fit (see the \textsc{mpfitexpr} documentation). These are lower than 0.5\,\AA\, on average for the central wavelengths and widths.\\

\noindent In order to map the velocity dispersion of the gas, the intrinsic line widths were computed after we removed the instrumental profile inferred from the sky lines (Sect.\,\ref{S_datared}). The flux maps were obtained directly from the measurements of line intensity on a spaxel-by-spaxel basis.\\
The line maps are shown in Figures  \ref{Fig_maps} and  \ref{Fig_map_bpt}. H$\beta$ and [O\,III] share the same kinematics, as mentioned above.


\section{Main observational results}
\label{S_results}

\begin{figure*}
\centering
\includegraphics[clip=true,width=0.975\textwidth]{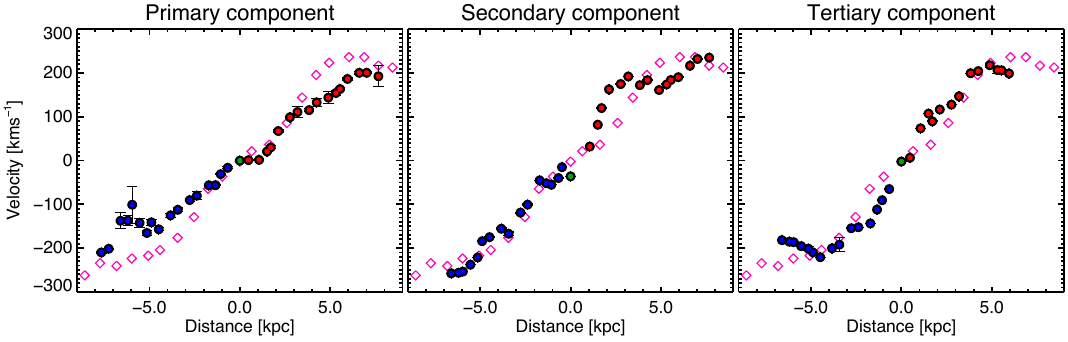}
\caption{Position-velocity (P-V)  curves of the different components (labelled at the top) that probe the ionised gas in Mrk\,1498. The curves were obtained considering a 0$\farcs$8 width pseudo-slit aligned with the major rotation axis (i.e. 148$^{\circ}$; Table\,\ref{T_kin}). There is a small offset between the kinematic centre of the components (filled green circles), i.e.$<$0$\farcs$8. The radius was therefore calculated as the distance from the photometric centre (continuum peak). For comparison,  the pink symbols indicate the P-V curves from \citet{Keel2012} with PA\,=\,149$^{\circ}$.}
\label{Fig_PV_PS_MajAx}
\end{figure*} 

The ionised gas probed by the [O\,III] emission lines is mostly distributed in the north-south direction. This is in contrast to the orientation of the elliptical host galaxy, PA\,=\,102$^{\circ}$ \citep{Vennik2000}, which was calculated based on B and R photometry obtained with CAFOS, which is mounted at the Calar Alto 2.2m telescope (spatial resolution of 0\farcs4/pixel).
\noindent  As mentioned in Sect.\,\ref{S_analysis}, the primary (broadest) component is the most extended and nearly covers the entire MEGARA FoV. \\
 The ionised gas probed by the two narrower components (secondary and tertiary) is found to have a similar extension (up to $\sim$\,6$\farcs$0, i.e. 7.1\,kpc). The main difference is that part of the secondary component emission is found to be spatially coincident with the direction of the radio jet towards the north-east (PA$_{\rm jet}$\,=\,45$^{\circ}$; Table\,\ref{T_kin}; Fig.\,\ref{Fig_maps}, second row). \\ 
For all the three components, the faint emission probed by H$\beta$ has the same morphology as [O\,III], but at a lower S/N. \\

\noindent We summarise below the main results for the three components we used to model the emission lines. We focus on kinematics (Sect.\,\ref{SS_kinematics}) and flux (Sect.\,\ref{SS_fluxes}). In Table\,\ref{T_kin} we indicate the main kinematic properties.


\begin{table}
\caption[]{Main kinematic properties of the different components of Mrk\,1498}
\begin{center}
\tiny{\begin{tabular}{l  c c c c}
\hline \hline
Component &  PA    & $\Delta$\,V & $\sigma_{\rm c}$ & $\sigma_{\rm avg}$\\
          & degree & \kms & \kms & \kms\\
\hline
Primary          & 148 & 231\,$\pm$\,12 & 191\,$\pm$\,34 & 96\,$\pm$\,24 \\
Secondary        & 155 & 247\,$\pm$\,5  &   22\,$\pm$\,6 & 25\,$\pm$\,9 \\
Tertiary         & 150 & 228\,$\pm$\,4  &  61\,$\pm$\,10 & 37\,$\pm$\,12 \\
 \hline
\end{tabular}
\label{T_kin}}
\end{center} 
\tiny{Notes. --- PA is the position angle of the kinematic major  axis; $\Delta$\,V is the observed peak-to-peak semi-amplitude of the velocity field; $\sigma_{\rm c}$ ($\sigma_{\rm avg}$) is the central velocity dispersion within (excluding) the nuclear region (see Sect.\,\ref{SS_kinematics}). 
The quoted uncertainties for the velocity dispersion measurements are one standard deviation.} 
\end{table}


\subsection{Kinematics}
\label{SS_kinematics}

\noindent  The velocity fields for the three components show the same overall behaviour with positive (negative) velocities  to the north-west (south-east). The components share a  position angle that is similar to that of the maximum velocity gradient and amplitudes (see the dashed lines in all the spectral maps and  Table\,\ref{T_kin}).\\
The main difference of the velocity maps is that the broadest (primary) component shows a blob at a distance of 1$\farcs$8 (2.2\,kpc) towards the south-west in the direction of the radio jet. This blob is observed with positive velocities ($\sim$\,75\,\nskms). At the same spatial position, the  $\sigma$-map shows values up to $\sim$\,175\,\nskms, which is higher than what is seen elsewhere (96\,$\pm$\,24\,\nskms; $\sigma_{\rm avg}$ in  Table\,\ref{T_kin}).\\

\noindent The primary component is about three times broader overall than the secondary and tertiary (excluding the nuclear PSF region). The width ratio of the two narrow components is 0.6 on average (the secondary component is narrowest).\\
The $\sigma$-map of the secondary component is rather homogeneous, with an average velocity dispersion in the central region ($\sigma_{\rm c}$) similar to that measured elsewhere (Table\,\ref{T_kin}). Only small deviations exist to the south-east and west of the nucleus at $\sim$\,40\,\nskms. These deviations lack a counterparts in the velocity or flux intensity maps. \\
\noindent The $\sigma$-maps of the primary and tertiary components show a similar behaviour. Specifically, they are both enhanced in the same directions: in the direction of the radio jet, and a direction that is almost perpendicular to it (PA $\sim$ -70$^{\circ}$). At these locations, the velocity dispersion is about 160\,-\,220\,\kms and 60\,-\,80\,\nskms for the primary and tertiary components, respectively. \\ 
None of the velocity dispersion maps is therefore as expected in the case of a regular rotating disc (see e.g. \citealt{Flores2006}). More specifically, the velocity dispersion maps are not centrally peaked. They are either rather flat (secondary component) or have some level of distortions (primary and tertiary components). This suggests the presence of ionised gas discs with different levels of perturbations (see Sect.\,\ref{SS_kinematicfeatures}).


\subsection{Emission line fluxes and line ratios}
\label{SS_fluxes}

We  describe the general behaviour of the flux-intensity of the [O\,III] line observed at higher S/N with respect to H$\beta$ below (as mentioned earlier). \\
\noindent The emission in the inner R\,$\sim$\,2\arcsec \ is not centrally peaked, but elongated (north-south)  for all three components. The elongation is strongest for the secondary component. The eccentricity of the ellipses contains 30\,$\%$ of the flux peak and
is 0.5 (Fig.\,\ref{Fig_maps}, middle row, third panel). The eccentricity is 0.6 and 0.8 for the primary and tertiary components, respectively.\\
This elongated morphological feature might be related to unresolved nuclear motions that cannot be distinguished at the spatial resolution of MEGARA (see Sect.\,\ref{SS_comparison}). \\

\noindent Up to large galactocentric distances ($\sim$\,6\,kpc), the evident features are gas clumps that are also visible  in the \textit{HST} image (see Fig.\,\ref{Fig_HST}). These clump-like features show log[O\,III]$\lambda$5007/H$\beta$  lower than 1.0 (0.90\,$\pm$\,0.08, on average) for the primary (most extended) component, as shown in Fig.\,\ref{Fig_map_bpt} (left). Elsewhere, the line ratio is 1.1\,$\pm$\,0.09  (see Fig.\,\ref{Fig_map_bpt}). \\
These clumps do not produce perturbations of the overall velocity dispersion maps or velocity field (see Fig.\,\ref{Fig_maps}). 


\section{Discussion}
\label{S_discussion}

In Sect.\,\ref{SS_comparison}, we compared the results from current MEGARA IFS data and previous works. The discussion in Sections \ref{SS_kinematicfeatures} and \ref{SS_clumps} is dedicated to kinematic and flux features, respectively, and to their possible connection with a putative outflow. In Sect.\,\ref{SS_jet} we complete the considerations of the radio jet activity periods (from previous works) and the optical IFS feature analysed here.\\
Figure\,\ref{Fig_PV_PS_MajAx}   presents the position-velocity plots along the major kinematic axis (see Table\,\ref{T_kin}) for all three components. Figure\,\ref{Fig_cartoon} is a schematic illustration of the observed features of Mrk\,1498 from previous multi-wavelength studies and the current work.\\


\subsection{Comparison with previous works based on optical data}
\label{SS_comparison}

\begin{figure}
\centering
\includegraphics[clip=true,width=0.475\textwidth]{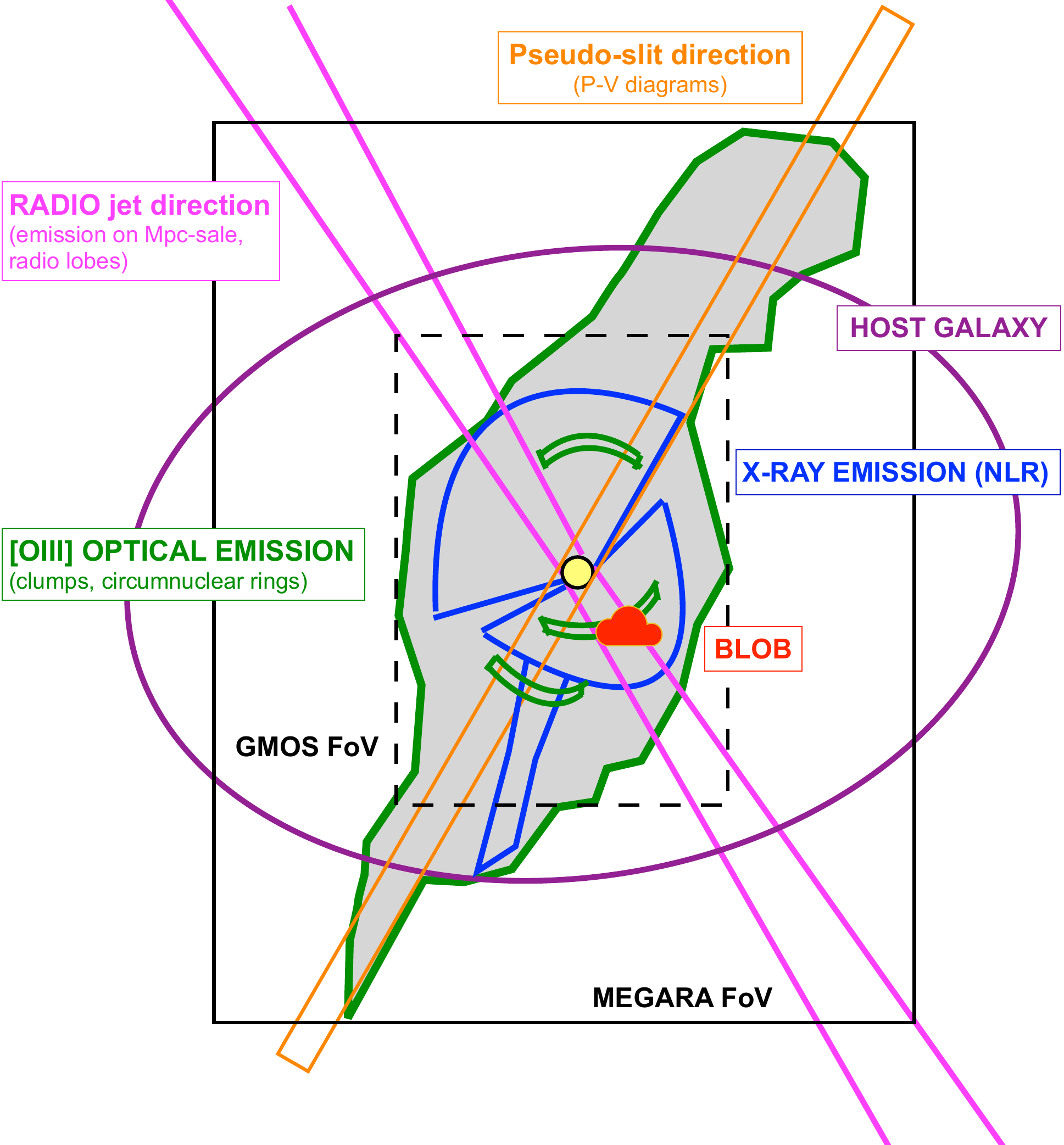}
\caption{Cartoon illustrating the different features of Mrk\,1498 from previous multi-wavelength studies and the current work (see text for details). The MEGARA and GMOS fields of view are also indicated. The yellow circle marks the nucleus.}
\label{Fig_cartoon}
\end{figure}

The study of the optical properties of both the AGN and the ionised gas emission in Mrk\,1498 is mostly based on slit spectroscopy with a low spectral resolution \citep{Keel2012, Keel2015, LHG2019}, except for the unique IFS-based work by \citet{Keel2017}. The latter was based on data from the Gemini Multiple-Object Spectrometer (GMOS) that covered a region of 3$\farcs$5\,$\times$\,5$\farcs$0 and had a image quality of 0$\farcs$4 in FWHM at a resolution of 0.5\,\AA\, per pixel. The GEMINI data set allowed us to map the rings close to the nuclear region that is observed in the Hubble image (Fig.\,\ref{Fig_cartoon}), but not the whole extended emission on kiloparsec scales that is covered by our MEGARA observation (Figures \ref{Fig_HST} and \ref{Fig_cartoon}). \\

\noindent The nature of the AGN in the nucleus of Mrk\,1498 has been debated in the last years (see \citealt{LHG2019}). At optical wavelengths, the gas surrounding the nucleus is ionised by the AGN radiation field, as indicated by the results from the analysis of slit-spectroscopy data from the Lick and San Pedro Martir observatories by \citet{Keel2012} and \citet{LHG2019} at a spectral resolution of 4 and 2.5 \AA\,per pixel, respectively. This AGN-ionisation scenario is mostly supported by means of standard Baldwin, Phillips $\&$ Terlevich (BPT) diagrams  \citep{Baldwin1981}. These are empirically derived diagrams based on optical emission line ratios (selected to be unaffected by reddening) that help us to distinguish different ionising mechanisms.\\
With the current MEGARA data that only cover H$\beta$ and [O\,III] lines (see Sect.\,\ref{S_analysis}), we cannot use the BPT diagrams to properly study the possible ionisation mechanism of the different components found in our analysis. The values of log ([O\,III]/H$\beta$) still only provide indications for an investigation of the dominant ionisation mechanism. Without the clumpy features mentioned in Sect.\,\ref{SS_fluxes}, the typical values of  log([O\,III]/H$\beta$)  are $>$\,0.9 (Fig.\,\ref{Fig_map_bpt}). Considering these typical values and the dividing lines that distinguish ionization from AGN and star formation by  \citet{Kauffmann2003} and \citet{Kewley2006}, for the line ratios for the two narrow components suggest that ionisation from AGN is the dominant mechanism. 
For the primary component, the line ratios are not higher than the  dividing lines above, with log([O\,III]/H$\beta$) $\sim$ 0.9-1.1. It is therefore essential to measure other standard line ratios to determine the dominant ionisation mechanism.\\
The clumpy structures characterised by log\,([O\,III]/H$\beta$)\,$<$\,1 are discussed in Sect.\,\ref{SS_clumps}.\\

\noindent  All the works based on optical slit spectroscopy indicate that the ionised gas in Mrk\,1498 has a kinematics that is dominated by rotation, with  extreme velocities of up to 300\,\nskms, as measured at large scales (up to R\,$\sim$\,25\,kpc). In addition  to the rotation component, previous works indicated multiple components and distortions in the velocity field \citep{Keel2012, Keel2015}. \\

\noindent  As we mentioned in Sect.\,\ref{SS_kinematics}, the kinematics of all the three components shows a clear velocity gradient in the north-south direction. From the maps of all the components, we extracted the kinematic values in a 0\farcs8 pseudo-slit along  the maximum velocity gradient\footnote{The difference in PA of the different components is not significant (i.e. smaller than 10$^{\circ}$). Therefore, we preferred to use a common PA.}. The PA is 148$^{\circ}$.    The position-velocity plots in Fig.\,\ref{Fig_PV_PS_MajAx} were obtained from the current MEGARA IFS data in order to compare our kinematic results with those from previous optical spectroscopy \citep{Keel2012}. For all the three components, the velocity curves (P-V curves, Fig.\,\ref{Fig_PV_PS_MajAx}) do not reach the plateau within the spatial scales sampled by MEGARA and show some irregularities at R\,$>$\,3\,kpc. For the secondary component, some irregularities are already evident at negative velocities at R\,$>$\,2\,kpc. These could be due to local non-ordered or non-circular motions at large galactocentric distances produced by instabilities from the previous merger episode or non-circular large-scale motions. The velocity amplitude and P-V curve (Table \ref{T_kin} and Fig.\,\ref{Fig_PV_PS_MajAx}) from current MEGARA maps agree with those obtained by \citet{Keel2012} along the slit location with PA of 149$^{\circ}$ (their Fig.\,12). Nevertheless, the measurements do not agree with the value of the velocity amplitude ($\pm$\,700\,\nskms) reported by \citet{Keel2017} using GMOS IFS-data (their Fig.\,9). We find no indication of high velocities like this at the corresponding spatial locations. The same authors also reported H$\beta$ and [O\,III] line profiles are narrow with red wings in the south, and about 2$\farcs$0 east of the nucleus. As mentioned in Sect.\,\ref{S_analysis}, narrow profiles are rare in MGARA data. The MEGARA spectral resolution allows us to reveal and model complex line profiles on spaxel-by-spaxels basis that are mostly characterised by double peaks and broad wings (see Fig.\,\ref{Fig_ajustes}).\\

\noindent  As mentioned in Sect.\,\ref{S_intro}, the optical \textit{HST}-images of the [O\,III] emission analysed by \citet{Keel2012,Keel2015} show two remarkable features. On the one hand, they show multiple circumnuclear rings of  ionised gas at radii of 0$\farcs$5\,-\,1$\farcs$6  (0.6\,-\,1.8\,kpc). On the other hand, relatively bright bubble-like structures that extend out to $\sim$\,10$\arcsec$ ($\sim$\,10\,kpc) to the north-east and south-west are present.
For MEGARA data, all the three components show an elongated flux distribution in the innermost region (up to R\,$\sim$\,3$\arcsec$; Fig.\,\ref{Fig_maps} and Sect.\,\ref{SS_fluxes}). This could be related to the ring-like morphology seen in \textit{HST} data but seen at lower spatial resolution with MEGARA-IFS. Similarly, for the large scale emission, two bubble-like features departing from the nucleus are well visible also in MEGARA data, their clumpiness is better seen in the log[O\,III]/H$\beta$ maps (Fig.\,\ref{Fig_map_bpt}).


\subsection{Kinematic features}
\label{SS_kinematicfeatures}

\noindent The three velocity components have similar peak-to-peak velocities overall, but different average and central values of the velocity dispersion. Based on their spatially resolved behaviour, none of the components would trace a regular rotating disc according to the classification by \citet{Flores2006}. Following this classification, the kinematics maps (velocity and velocity dispersion) of the primary and tertiary components might both be classified as  \lq perturbed discs\rq: Their velocity maps are fairly regular, but both deviate from the case of an ideal rotating disc. Although the primary and tertiary components share the same disc classification, we remark that the disc traced by the tertiary component is the more perturbed of the two, with a more disturbed velocity dispersion map. The gas probed by the secondary component follows the main rotational pattern with the lowest values of the velocity dispersion; the main features of its velocity dispersion maps suggest complex motions, as discussed below.\\

\noindent The primary component shows a higher velocity dispersion ($\sim$\,170\,\nskms) than average in the $\sigma$-map ($\sim$\,96\,\nskms; Table\,\ref{T_kin}) that is mostly spatially extended along the radio axis. At the same spatial location, but only towards the south-west, lies a blob with a velocity up to 100\,\kms. The velocity is not extreme, but is observed in the direction of the minimum velocity gradient (i.e. north-east/south-west) and in the direction of the jet propagation. \\

\noindent The high-$\sigma$ values of the primary component could be associated with shocks and induced turbulence due the passage of the jet through the plane of the galaxy. This is consistent with the elongation of the core of the radio jet in the north-east/south-west direction detected by \citet{LHG2019}. Moreover, taking the position angle of the jet into account (PA$_{\rm jet}$\,=\,45$^{\circ}$, Table\,\ref{T_properties}), we propose that the jet is inclined with respect to the galaxy plane because we also observe perturbations in the tertiary component in two different directions (in the direction of the jet and nearly perpendicular to it in a sort of boomerang-like shape).\\

\noindent Signatures of episodic AGN and jet activity have been found in other Seyfert galaxies (e.g. NGC\,2639; \citealt{Rao2023}). We explored a scenario in which the blob at positive velocity seen towards the  south-west would be caused by either present or past nuclear activity. For this purpose, we integrated in the nuclear region (i.e. within the PSF size; circles in all figures) the observed [O\,III] fluxes for the primary component  ($\sim$\,5$\times$10$^{41}$ erg/s) and derived the corresponding luminosity. Using the [O\,III] luminosity, we calculated a kinematically determined outflow size following \citet{Kim2023}. The size of the putative outflow is $\sim$\,2\,kpc,  similar to the distance of the blob from the nucleus ($\sim$\,2.3\,kpc). This similar spatial likeness might indicate that the current AGN activity launched the blob.\\
\noindent However, taking the typical velocity at line peak of the blob into account ($\sim$\,100\,\nskms; Fig.\,\ref{Fig_maps}), we should detect a velocity dispersion of more than 350\,\nskms, that is, much higher than observed in our MEGARA data. If the scaling between the maximum outflow velocity (V$_{max}$) and AGN bolometric luminosity\footnote{We derived the AGN bolometric luminosity ($\sim$\,10$^{45}$ erg/s) from the 2-10 KeV Xray luminosity (23.31, in logscale, \citealt{LHG2019}) assuming the relation by \citet{Marconi2004}.}
 (L$_{bol}$)  holds (Table\,1 and Fig.\,2 in \citealt{Fiore2017}), the blob should move at V$_{max}$\,$\sim$\,500\,\nskms.\\

\noindent It is commonly thought that the AGN phenomenon is a phase in the lifecycle of a galaxy in which the nuclear activity can be reactivated 10-100 times during its lifetime, with typical timescales of $\sim$\,0.5\,Myr \citep{Schawinski2015}. This timescale is much shorter than the dynamical time of blob ($\sim$\,20\,Myr), calculated as in \citet{PereiraSantaella2016}. This indicates that the last episode of AGN activity, the restarted one, is probably not the power mechanism that launched the blob in the putative outflow.\\

\noindent Taking into account all this evidence, we propose a scenario in which  the blob might be associated with a putative outflow powered by a previous more luminous AGN episode. In this scenario, the outflow would have  quietly expanded throughout the galaxy with low levels of turbulence. This result is consistent with the one reported in \citet{LHG2019}, who found that the currently active nucleus is able to photoionise the gas in its surrounding, but not the large-scale emission on kiloparsec scales.\\
Based on the observed kinematics, an alternative scenario is that the young jet has created a blob at its present position and was not pushed out from the nucleus. However, this is difficult to investigate without the knowledge of the other BPT line ratios (e.g. [N\,II]/H$\alpha$), and hence test shocks models.\\

\noindent To summarise, a young jet is present in the nucleus of Mrk\,1498, as shown by a previous analysis of radio data \citep{Bruni2019}. However, the blob is probably not associated with the recent (restarted) activity, but with the already-faded previous nuclear activity.


\subsection{Flux features (clumps)}
\label{SS_clumps}

Similarly as in Sect.\,\ref{SS_kinematicfeatures} for the blob, we explored a scenario in which the clumps are indications of either present or previous nuclear activity. 

\noindent  The total extension of the clumpy structure is $\sim$\,6\,kpc in radius, and the values of the velocity dispersion of the gas clumps are between 20 and 110 \kms, without a clear dependence on the distance from the nucleus. The velocities are about 50-220\,\nskms. This implies that the dynamical times required for the gas to reach its current locations are 10-40 Myr (see also Sect.\,\ref{SS_kinematicfeatures}). The dynamical time of clumps is much longer than the typically time of an AGN duty cycle, indicating that the last episode of AGN activity, the restarted one, is probably not the power mechanism that launched the clumps.

\noindent When we assume that the size of the putative outflow is  $\sim$\,2\,kpc (Sect.\,\ref{SS_kinematicfeatures}) and that the clumps are seen up to $\sim$\,6\,kpc, these clumps are likely associated with  an AGN-driven outflow from a previous more luminous AGN episode or were launched by the actual restarting activity with an additional energy injection by the radio jet.\\

\noindent If the clumps were launched by the faded AGN activity, the putative outflow would be classified as  fossil. These outflows persist for an order of magnitude longer than the AGN episode driving it (see \citealt{Zubovas2020}). Other similar cases in the literature are PDS\,456 and IRAS F11119+3257 \citep{Zubovas2020}. \\
Mrk\,1498 shows a GPS in the centre \citep{LHG2019, Bruni2019}. If the outflow were associated with the current restarting young radio activity, the kinematics would be more extreme (large [O\,III] widths) and the ionised gas morphology would be more disturbed, as proposed by \citet{Kukreti2023}.\\
An alternative scenario is that the clumps are simply H\,II regions that follow the main rotation pattern of the ionised gas in the galaxy due to their low values of log[O\,III]/H$\beta$. \\

\noindent To summarise, similarly to the case described in Sect.\,\ref{SS_kinematicfeatures}, the crossing timescale of the clumps is longer than what is expected for a typical AGN active episode, which suggests that the large-scale outflow is unrelated to the current (restarted) AGN phase. The different kinematics, extension, and morphology of the blob (see Sect.\,\ref{SS_kinematicfeatures}) and the clumps suggest that the features are associated with different flickering AGN episodes in the past that are unrelated to the current young radio phase in the nucleus.

\subsection{Considerations about the jet activity periods}
\label{SS_jet}
The two distinct radio activity periods deduced for this source allow us to consider the jet activity period and its possible feedback on the host galaxy. The megaparsec-scale lobes are typically formed in a time span of $\sim$\,100 Myr (e.g. \citealt{Orru2014}), indicating that the jet (thus the AGN) remained active for a similar amount of time. The new radio phase found in the core, spotted through the presence of a peaked radio spectrum and confirmed by VLBI observations at parsec-scale resolution, typically has an age of a few thousand years and a linear size of $\sim$\,1 kpc \citep{2021A&ARv..29....3O}. The recent release of the Very Large Array Sky Survey (VLASS) survey images at 3 GHz (3" resolution; \citealt{2020PASP..132c5001L}) allows us to place a further constraint on the size of the recently launched jet. The source core shows a projected linear size of $\sim$0$\farcs$6  (deconvolved), corresponding to $\sim$\,600 pc at the redshift of the source, in agreement with the expected one for young radio sources. The reactivation of the jet should thus have occurred some $>600 \rm{pc}/c\sim2000$ years ago. The apparent discontinuity of $\sim$\,50 kpc between this new phase and the inner edge of the radio emission that is connected with the lobes \citep{Bruni2021} is thus most probably due to the quiescent period between the two jet activity periods, which could have lasted $>50\rm{kpc}/c\sim200\rm{kyr}$. The blob discussed in this work lies in the gap region between the previous and ongoing radio phases, where the jet that formed the megaparsec-scale lobes has deposited energy for some million years.\\
The AGN flickering may imply changes in the accretion state or AGN power. These are not taken into account in the current discussion because the current data set prevents us from developing a customised and detailed modelling of the full AGN history as was done in  \citet{Zubovas2022}. 

\section{Conclusions}
\label{S_conclusions}

\noindent On the basis of optical MEGARA IFS data, we have studied the properties of the ionised gas component in the Seyfert\,1.9  Mrk\,1498 after a multiple component fitting, using the H$\beta$ and [O\,III] emission lines as tracers.\\
The conclusions of this study are summarised below. 

\begin{enumerate}

\item \textit{Multiple component fitting.} For the first time, we were able based on R\,$\sim$\,10,000 optical IFS MEGARA data to disentangle three kinematic components and map their properties. This represents an improvement with respect to all the previous works that were based on optical spectroscopic data, which only studied one component with only weak detections of broad line-wings.\\

\item \textit{Kinematics.} All the three components show an overall blue-to-red velocity gradient, with similar peak-to-peak velocities, but a different velocity dispersion in the centre and on average. The visual inspection of the maps of two components highlights high velocity dispersion values in the direction of the radio jet. We kinematically classified the ionised gas probed by primary and tertiary components as  \lq perturbed discs\rq \ \citep{Flores2006}. The disc traced by the tertiary component is more perturbed than that of the primary component. The secondary component is likely associated with flux clumps in a disc with low levels of turbulence.\\

\item \textit{Ionisation mechanisms.} There is no clear trend of the [O\,III]/H$\beta$ ratios on the distance overall. However, our MEGARA maps clearly reveal clumps with line ratios lower than unity in the direction of the maximum velocity gradient. The observed line ratios indicate possible ionisation from AGN or shocks everywhere else.\\

\item \textit{Kinematic features.} Towards the south-west at a galactocentric distance of $\sim$\,2.3\,kpc, we observe a blob with a velocity up to 100\,\nskms and a high velocity dispersion ($\sim$\,170\,\nskms) that is spatially coincident with the direction of the radio jet. Taking into account its kinematic, dynamical time ($\sim$\,20\,Myr), and spatial location, we proposed a scenario in which the blob is powered by a previous more luminous AGN episode with possible additional energy input from the radio jet. We suggest that the jet is likely inclined with respect to the galaxy plane, as we also observe a high velocity dispersion in the tertiary component in two different directions (in the direction of the jet and nearly perpendicular to it in a sort of boomerang-like shape). \\

\item \textit{The clumpy structure.} The dynamical time of the clumps (10-40\,Myr) is much longer that the typically time of the AGN duty cycle (0.5\,Myr). The clumps are detected up to 6\,kpc, which is much farther than the expected radius of a putative outflow launched from the current restarting activity. We proposed two possible scenarios in which the clumps are either an indication of the faded AGN activity (a fossil outflow) or are simply HII regions that follow the main rotation pattern of the ionised gas in the galaxy due to their low values of log[O\,III]/H$\beta$.\\

\item \textit{Radio jet activity periods.} Radio lobes on megaparsec scales similar to those found in Mrk 1498 typically form in a time span of $\sim$100 Myr, whereas the new radio phase found in the core typically has an age of a few thousand years. The reactivation of the jet should thus have occurred  $\sim$\,2000 years ago. The apparent discontinuity is thus most probably due to a quiescent period between the two jet activity periods.

\end{enumerate}

\noindent In summary, based on the optical and radio activity, we propose that two different ionised gas features are observed over the radio AGN lifecycle of Mrk\,1498. On the one hand, we observe at radio frequencies old radio lobes (from the faded AGN) and a young GPS nucleus (from the current restarted AGN). On the other hand, we observe clumps and a blob at optical wavelengths that are likely associated with the past AGN activity (not the current activity) and are likely fossil-outflows related to two different episodes of the flickering AGN activity. When we take the time gap among radio activities and the jet-size evolution into account, the blob discussed in this work could have been launched between the previous and the ongoing radio phases.


\begin{acknowledgements}

SC, IM, and JM  acknowledge financial support from the Severo Ochoa grant CEX2021-001131-S funded by MCIN/AEI/ 10.13039/501100011033. These authors are also supported by the Spanish Ministry of Science, Innovation y University (MCIU) under grants PID2019-106027GB-C41 and PID2022-140871NB-C21.  \\
LHG acknowledges financial support from FONDECYT Iniciacion 11241477, and ANID Millennium Science Initiative ICN12$\_$009.\\
GB acknowledges financial support from the Bando Ricerca Fondamentale INAF 2023, for the project: \textit{The GRACE project: high-energy giant radio galaxies and their duty cycle}.  GB acknowledges financial support for the GRACE project, selected via the Open Space Innovation Platform (\url{https://ideas.esa.int}) as a Co-Sponsored Research Agreement and carried out under the Discovery programme of, and funded by, the European Space Agency (agreement No. 4000142106/23/NL/MGu/my).\\
FP acknowledges financial support from the Bando Ricerca Fondamentale INAF 2023: Exploring the origin of radio emission in Radio Quiet AGN. \\

\noindent This research has made use of the NASA/IPAC Extragalactic Database (NED), which is operated by the Jet Propulsion Laboratory, California Institute of Technology, under contract with the National Aeronautics and Space Administration. We acknowledge the usage of the HyperLeda database (\url{http://leda.univ-lyon1.fr}).
\end{acknowledgements}


\bibliographystyle{aa} 
\bibliography{Bibliography.bib}

\end{document}